\documentclass[conference]{IEEEtran}
\IEEEoverridecommandlockouts
\usepackage{cite}
\usepackage{amsmath,amssymb,amsfonts}
\usepackage{algorithmic}
\usepackage{graphicx}
\usepackage{textcomp}
\usepackage{xcolor}
\def\BibTeX{{\rm B\kern-.05em{\sc i\kern-.025em b}\kern-.08em
    T\kern-.1667em\lower.7ex\hbox{E}\kern-.125emX}}
\begin{document}

\title{Exploring the Integration of Extended Reality and Artificial Intelligence (AI) for Remote STEM Education and Assessment\\

}

\author{\IEEEauthorblockN{Shadeeb Hossain}
\IEEEauthorblockA{\textit{Research Division} \\
\textit{Shadeeb Engineering Lab}\\
New York, USA \\
ORCID : 0000-0002-5224-7684\\
shadeeb@shadeebengineeringlab.com}
\and
\IEEEauthorblockN{Natalie Sommer}
\IEEEauthorblockA{\textit{College of Engineering and Information Systems} \\
\textit{DeVry University}\\
New York, USA\\
nsommer@devry.edu}
\and
\IEEEauthorblockN{Neda Adib}
\IEEEauthorblockA{\textit{College of Engineering and Information Systems} \\
\textit{DeVry University}\\
California, USA\\
Neda.Adib@devry.edu}
}

\maketitle

\begin{abstract}
This paper presents a dynamic gamification architecture for an Extended Reality – Artificial Intelligence (XR-AI) virtual training environment designed to enhance STEM education through immersive adaptive, and kinesthetic learning. The proposed system can be introduced in four phases: Introduction Phase, Component Development Phase, Fault Introduction and Correction Phase and Generative AI-XR scenarios Phase. Security and privacy are discussed via a defense-in-depth  approach spanning client, middleware, and backend layers, incorporating AES 256 encryption, multi-factor authentication, role-based access control (RBAC) and GDPR/FERPA compliance. Risks such as sensor exploitation, perceptual manipulation, and virtual physical harm are identified, with mitigation strategies embedded at the design stage. Potential barriers to large scale adoption-including technical complexity, cost of deployment, and need for cybersecurity expertise are discussed. 
\end{abstract}


\section{Introduction}
According to data from the National Centre for Education Statistics, approximately 9.4 million undergraduate students, about $61\%$ of total enrollment in Fall 2021, were enrolled in at least one distant learning course. However, STEM discipline remains underrepresented in online learning due to the inherent challenge of delivery hands-on training and laboratory experiences virtually. Traditional e-learning platforms lack interactivity, spatial visualization, and real-time feedback essential for STEM pedagogy, such as conducting experiments interpreting instrumentation, or interacting with complex simulation [1-4].

\setlength{\parskip}{1em}
Emerging immersive technologies – such as extended reality (XR), encompassing augmented reality (AR) , virtual reality (VR) and mixed reality (MR) offer a transformative pathway to enhance remote STEM education. XR enables high fidelity simulation of laboratory environments, interactive procedural training , and immersive visualization of abstract concepts, thereby addressing critical barriers in online STEM instruction [5-7].

\setlength{\parskip}{1em}
Extended reality (XR) encompasses a range of immersive technologies, including Virtual Reality(VR) , Augmented Reality (AR) and Mixed Reality (MR) [8]. Virtual reality enables users to interact with a fully simulated digital environment using specialized interfaces such as head-mounted displays and haptic devices [9]. Augmented Reality (AR) on the other hand, integrates computer-generated content with the physical world in real time, enhancing the user’s perception of and interaction with real world contexts [10,11]. Augmented Virtuality (AV) refers to the incorporation of real-world objects into a primarily virtual environment, forming part of the Reality-Virtuality continuum [12].  These immersive technologies collectively support experiential and spatial learning, making them particularly promising for applications in STEM education and remote training environments. 

\setlength{\parskip}{1em}
The Sensorama Simulator (1962) represents one of the earliest patented efforts to develop immersive systems capable of delivering multisensory virtual experiences [13]. Originally designed to address the pedagogical challenge of conveying complex subjects to large, diverse student populations with limited specialized teaching staff. Sensorama introduced a novel integration of sensory modalities – including olfactory stimuli, haptic feedback, and directional audio. These enhancements overcame the limitation of conventional visual aids, which lacked realism and sensory engagement. Sensorama’s design laid foundational principles for extended reality (XR) system, now positioned to transform STEM education through immersive, scalable, and resource-efficient remote training environment. 
Over the decades, advancements in computing power, graphics rendering, and human-computer interaction have enabled XR to evolve from static simulations into highly interactive, networked, and AI assisted environments. Recent developments in XR for education, along with growing research adoption, have created opportunities for cost effective hybrid training platforms. 

\setlength{\parskip}{1em}
Liarokapis et al. (2024), introduced  “Extended Reality for Education (XR4ED)”, a platform enabling educators to create XR-based teaching resources through a marketplace of 3D avatars and virtual environments without requiring extensive programming expertise [14]. In its pilot phase, the architecture incorporated an intelligent non-player AR character capable of interacting with learners and answering their queries. Future iterations aim at enhancing its avatar communication and improving synchronization between body coordination and speech. 
Berguld (2024)  investigated professional case-based learning scenarios to evaluate design approaches and user experience [15]. Four representative cases were analyzed : (i) AR demonstration, (ii) AR Development and Implementation- focused on introducing the HoloLens to prospective users , (iii) VR Assembly and Testing- supporting product assembly validation and installer training, and (iv) VR Assembly, Development and Testing- providing a user friendly programming interface for importing external CAD files, defining workflow logic, and enabling assembling and dissembling tasks with minimum instructor intervention . 

\setlength{\parskip}{1em}
Gu et al. (2024) proposed a novel XR environment tailored for Artificial Intelligence (AI) education [16]. Their framework leverages the spatial affordances of 3D VR to create immersive learning environment for understanding complex topics such as neural network architecture. Emphasis was placed on utilizing and leveraging the 3D VR spatial properties to create an immersive learning experience and engagement. The study also focused on optimizing interaction paradigms to help stimulate natural interactions and visualization techniques. 

\setlength{\parskip}{1em}
These studies collectively underscore the increasing adoption of XR-based educational platforms emphasizing accessibility, immersion and domain-specific adaptability. DeVry University, a predominantly online and hybrid higher educational institution, serves a geographically dispersed student population. In STEM curriculum, peer collaboration, personalized learning pathways, and simulated training for engineering case studies are critical to enhancing student engagement, retention and satisfaction.

\setlength{\parskip}{1em}
This paper proposes an XR gamification architecture with a specific focus on engineering education. The structure of the paper is as follows: (i) Gamification in Extended Reality – discusses literature review and analysis of gamification strategies applied to improve learning outcomes, (ii) Proposed Architecture for XR Gamification at DeVry University – a phased integration strategy for seamless incorporation into the curriculum, (iii) System Architecture for XR-AI Unity platform – it discusses the detailed description of the client layer, middleware layer and backend layer (iv) Privacy and Security- considerations and mitigation strategies for privacy and security aspects for the project (v) Summary and Future Directions- it discusses the comprehensive overview of the proposed XR-AI system and its scalability to other higher education and  K-12 contexts. 
The proposed hybrid model enables remote learners to engage in kinesthetic training, interaction, troubleshooting exercises, and complex engineering case studies previously limited to supervised on-site laboratories. The architecture employs Unity  for immersive environment development , Photon Enterprise Cloud for real-time scalability and multi-user interactions, and AI-driven analytics for adaptive performance assessment, engagement tracking, and dynamic  tailoring for  subsequent learning modules.

\section{Gamification in Extended Reality (XR) }

A study by Dong et al. (2024) focused on combining core elements of student-centered learning and gamification within an extended reality (XR) educational platform [17]. The researchers developed a virtual simulation-based electric power training system, selected for its complex characteristics. The gamified XR training model was found to be both effective and satisfactory among participants, highlighting the potential of an immersive gamified training environment. 

\setlength{\parskip}{1em}
Another study by Danielle (2025) explored the integration of: (i) gamification, (ii) generative AI, and (iii) Extended Reality (XR) to address three common challenges in online learning such as : (i) student dropout rates, (ii) learner engagement, and (iii) interaction [18]. The study yielded the following insights: (i) prompt-based generation of multimedia content (enabled by generative AI) enhanced instructional practices, (ii) 3D models (via XR) improved learner understanding and satisfaction, (iii) A card-based gamification system, which assigned personalized challenges and provided individual feedback, significantly boosted student motivation and conceptual understanding. 
Zikas et al. (2016) also examined the role of Mixed Reality (MR) digital games in the future of education [19] . Their case study involved a primary school history class that used Mixed Reality Serious Games and Gamification (MRSG), incorporating gesture- and game- based  learning. A desktop-based holographic application  with Meta-AR glasses and integrated game shells was employed. The learners interacted through gesture-based controls to complete tasks, and the effectiveness of this approach was evaluated. 

\setlength{\parskip}{1em}
These studies collectively demonstrate the effectiveness of XR and gamification in enhancing learner experience, participation, and engagement in virtual or online educational environments. This evidence supports the potential adoption of similar gamified XR learning strategies at DeVry University (DU). As DU primarily operates as an online institution with campuses and students spread across various locations in the United States, these strategies could encourage collaboration across state lines- especially in group projects -while boosting overall engagement and learning outcomes.  
Extensive research has shown that gamification in education is most effective when tailored to individual learner’s needs [20]. Game elements must be customized to learner’s profile, and this is where the integration of generative AI and XR becomes particularly valuable. Together, they enable the creation of adaptive gamification XR learning environments, which can be implemented using two approaches: (i) static and (ii) dynamic adaptation. Fig.1 presents a schematic illustrating the differences between (a) static and (b) dynamic gamification process. 

\setlength{\parskip}{1em}

\begin{figure}[htbp]
\centering
\includegraphics[width=0.9\linewidth]{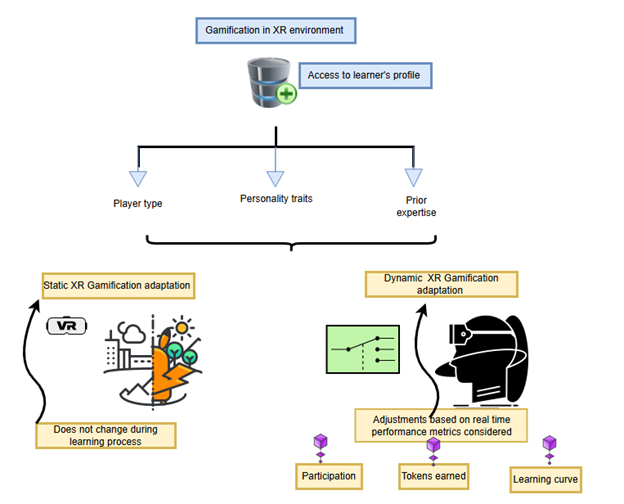} % scale down
\caption{Schematic illustrating the differences between (a) static and (b) dynamic gamification process.}
\label{fig:gamification}
\end{figure}

In static adaptation, the gamified environment is configured before the learning process begins, often based on user profiles. This approach employs  a generalized structure that does not change during the learning experience. Halifax et al. (2019) proposed that initial factors such as (i) player type, (ii) personality traits, and (iii) prior expertise can inform the allocation of appropriate gamified XR platforms [21]. 

\setlength{\parskip}{1em}
In contrast, dynamic adaptation modified the learning experience in real time, responding to the learner’s performance. Generative AI plays a critical role in facilitating these on-the-fly adjustments. Learners can set goals, and the system dynamically alters gaming elements-such as challenges, feedback, or complexity-based on real-time performance. The number of steps on difficulty levels required to achieve learning targets can be adjusted based on the learner’s prior knowledge and ability to incorporate feedback. Key parameters for dynamic adaptation may include: (i) participation or interaction level with the XR environment, (ii) scores or tokens earned during assessments, and (iii) individual learning curves.

\section{Proposed Architecture for XR Gamification in Engineering Course at DeVry University}
The XR gamification framework is particularly relevant for institutions such as DeVry University, where students are geographically dispersed. XR technology can enhance peer collaboration, support self-paced learning, and provide equitable access to engineering experiences that are traditionally constrained by physical infrastructure. It offers a powerful means of enriching electrical engineering education by simulating real -world scenarios that are otherwise difficult to replicate. A notable example is the use of XR  in safety training for handling high voltage equipment – an area where immersive, interactive environment can significantly improve learner understanding and preparedness. 

\setlength{\parskip}{1em}
Due to the complexity involved in developing an XR gamification system tailored to electrical engineering applications, the project is structured into multiple phases: (i) Introduction Phase, (ii) Component Development Phase (iii) Fault introduction and Correction Phase, (iv) Generative AI-XR scenarios phase. 

\setlength{\parskip}{1em}
The primary objective is to design a dynamic XR- based gamification system that adapts to each learner’s interaction within various scenarios. The system will consider factors such as tokens or  scores earned during testing stages, user behavior in each scenario, and the learning curve demonstrated through feedback integration across modules. This adaptive framework aims to provide a more personalized and engaging educational experience, better aligning with individual learning needs and professional engineering contexts. 

\setlength{\parskip}{1em}
\subsection{Introduction Phase}

In the Introduction Phase, the XR system with be developed using Unity XR within a virtual classroom environment. This phase aims to test the hypothesis that virtual collaboration is enhanced in the metaverse compared to traditional e-learning platforms. The effectiveness of the XR classroom will be evaluated using several student engagement metrics, including: (i) contribution to classroom discussions in the metaverse versus online platforms, (ii) student satisfaction, (iii) student attendance in the metaverse environment, (iv) test scores following metaverse-based classes utilizing a kinesthetic learning approach, compared to scores from students attending similar content delivered via conventional  online lectures. 
The rationale behind using XR to simulate a metaverse-style classroom for DeVry University lies in its ability to foster a more embodied and immersive learning environment. While remote learning offers the flexibility of eliminating commutes and enabling asynchronous access, the XR classroom adds  significant value by promoting peer interaction and enabling kinesthetic learning through simulated engineering models and troubleshooting of hypothetical technical faults. This immersive setting is expected to enhance both peer collaboration and cognitive presence, contributing  to improved comprehension and knowledge retention. By comparing key metrics- such as class discussion participation, student satisfaction, attendance rates, and post class test scores- the study aims to validate the hypothesis that a metaverse- based learning environment provides a more engaging and effective alternative to conventional e-learning platforms. 
Fig. 2 shows the Unity XR environment elements used to simulate a virtual classroom for DeVry University : (a) virtual campus imported from “Unity Assets”; (b) virtual table and chair setup; (c) virtual green board; (d) virtual laptop; and (e) assets listed used to create a realistic and interactive XR classroom environment.  For this initial proof-of-concept, the “School Asset” package available on Unity Assets was utilized. This package includes approximately 145 items -such as books, shelves, desks, projectors, monitors, and keyboards with a file size of 1.2 MB. 
The primary objective is to replicate a typical  DeVry University classroom, where avatars representing enrolled students can interact with peers and instructors in a lifelike virtual setting.  XR headsets can be provided to students, enabling them to connect remotely and participate in the metaverse -based learning environment, enhancing engagement and collaboration regardless of physical location.

\begin{figure}[htbp]
\centering
\includegraphics[width=0.9\linewidth]{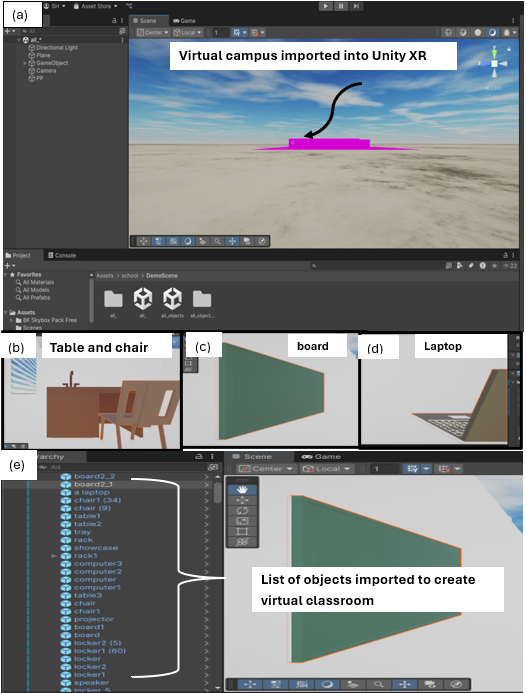} % scale down
\caption{Unity XR environment elements used in simulating the virtual classroom for DeVry University : (a) virtual campus imported from “Unity Assets”; (b) virtual table and chair setup; (c) virtual green board; (d) virtual laptop;  (e) assets listed used to create a realistic and interactive XR classroom environment 
}
\label{fig:XR Environment}
\end{figure}

\subsection{Component Development Phase}
The Component Development phase presents a significant challenge, as it involves the assembly of various electronic components within the XR environment. The initial step includes developing or importing 3D models of basic components such as resistors, capacitors, inductors , wires, power sources for more complex components, advanced 3D CAD models-such as printed circuit boards- can be sourced from platforms like Sketchfab. Fig. 3 illustrates a  schematic overview of the Component Development phase,  detailing each step of the process. 

\begin{figure}[htbp]
\centering
\includegraphics[width=0.9\linewidth]{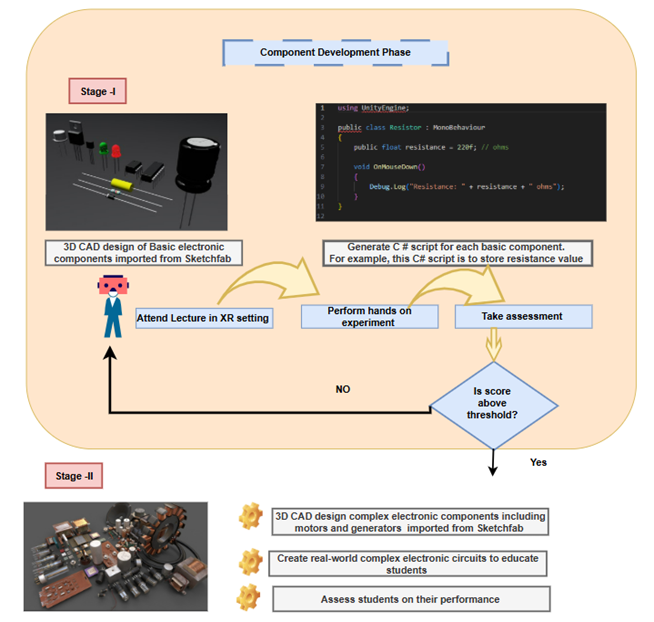} % scale down
\caption{Schematic of the Component Development phase with a detailed  process overview.  
}
\label{fig:Process Overview}
\end{figure}

\setlength{\parskip}{1em}
The Component Development phase can be divided into multiple stages, as shown in Fig. 3. In the initial stage, 3D computer aided design (CAD) models can be developed in Unity or imported from external sources into the XR environment. ChatGPT can also assist in generating relevant  scripts, as demonstrated in our previous work on the design and development of an RISC processor [22]. At DeVry University, multiple students enrolled in the same course can first attend a virtual lecture, followed by laboratory- based experiments to support kinesthetic learning. Students are then assessed to track their progress, with performance metrics stored in the Photon Cloud, as detailed in the subsequent section: System Architecture for proposed XR-AI Unity platform. Based on their performance, the system algorithm may allow students to progress to more advanced stages or repeat earlier ones to reinforce understanding. This approach also minimizes personal bias and reduces the impact of grading curves on student evaluations. 

\setlength{\parskip}{1em}
\subsection{Fault Introduction and Correction Phase }

One of the key advantages of Extended Reality (XR) in engineering and STEM education is its ability to simulate complex scenarios that are challenging to replicate in real-life settings. Notable examples include: (i) training medical first responders (MFR) through simulated casualty incidents, (ii) providing real-time clinical support training for deep space missions, and (iii) radiology education and immersive training environments [23-25]. XR has also been increasingly adopted in engineering education, particularly in context of Construction 4.0 and Fifth Industrial Revolution, where the emphasis is on training skilled professionals to operate advanced machinery and systems [26-28]. 

\setlength{\parskip}{1em}
Engineering case studies can be developed to train DeVry University engineering students on complex technical scenarios that are otherwise difficult to replicate in physical labs. For instance, transformer models can be designed using CAD software and imported into Unity for interactive simulation. After an initial instructional phase covering transformer operation and maintenance, various fault conditions can be introduced for experimental learning. These may include: (i) overheating, (ii) oil leakage , (iii) core lamination faults , and  (iv) insulation failure. The virtual scenarios can closely mirror real world industrial environments, enabling students to analyze troubleshoot, and respond to system failures in a risk free yet realistic settings.  

\setlength{\parskip}{1em}
\subsection{ Generative AI – XR Scenarios }

To enhance the effectiveness of such XR based training, AI -powered adaptive assessment modules can be used. The assessment modules can monitor and record the student decisions, performance and other metrics during each fault scenario, provide instant feedback, and log responses for performance evaluations. Based on the learner’s interaction, history and success rate, the system can dynamically adjust future scenarios-either increasing complexity or revisiting foundational concepts. This adaptive feedback mechanism ensures personalized learning pathways and thereby provides an alternate for cost-effective , efficient learning outcomes in engineering education. 	
Generative AI-XR scenarios can reduce the need for constant supervision by instructors, allowing  AI- driven prompts to generate C\#  that are integrated into Unity for replicating case studies tailored to students learning needs. These dynamically generated scenarios can simulate complex engineering problems, enhancing engagement and personalized learning within the XR environment. Several prediction factors, such as- time spent in simulation , preferred learning modalities, response time during assessments , and overall performance score can be analyzed to customize and optimize the AI generated content [29].  

\section{System Architecture for Proposed XR-AI Unity Platform }
The system architecture for the XR gamification in engineering courses is composed of three layers: (i) Client layer, (ii) Middleware layer, and (iii) Backend layer as shown in Fig. 3.  

\begin{figure}[htbp]
\centering
\includegraphics[width=0.9\linewidth]{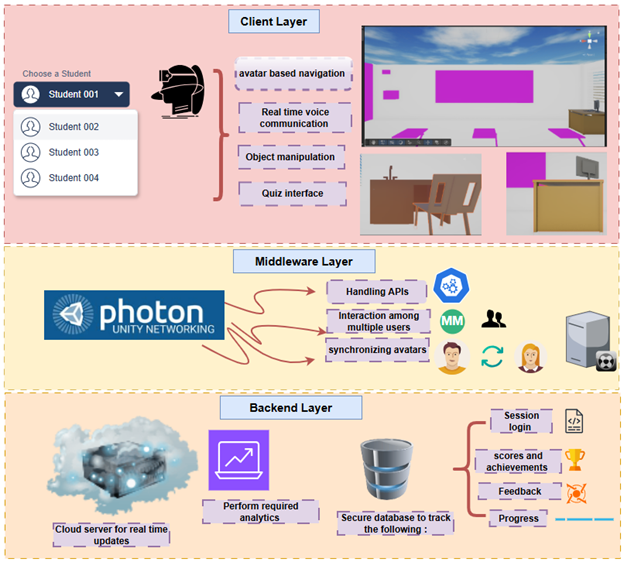} % scale down
\caption{The system architecture for the XR gamification platform in engineering courses is composed of three layers: (i) Client layer, (ii) Middleware layer, and (iii) Backend layer. 
}
\label{fig:Process Overview}
\end{figure}

\setlength{\parskip}{1em}
The client layer serves as the user interface  (UI) through which students interact with the virtual XR environment using VR headsets and optionally , haptic simulators.  This layer supports a range of functionalities, including real-time voice communication, object manipulation , and interactive quiz interfaces. A primary challenge at this level is designing a realistic and immersive metaverse experience that encourages student engagement and exploration of the platform’s features. Artificial Intelligence (AI) plays a key role in this layer by monitoring student satisfaction and tracking learning progress. Advanced AI algorithms can be employed to enhance user motivation by generating contextually relevant metaverse scenarios or by dynamically adapting learning content to optimize educational outcomes based on individual performance metrics.  

\setlength{\parskip}{1em}
The middleware layer is responsible for synchronizing avatars, managing multiplayer interactions, and handling API communications between clients and backend systems. Photon Unity Networking (PUN) is utilized to enable real time multiplayer functionality and to ensure low-latency communication through efficient transport layer. For the purposes of this application, Photon Fusion is recommended, as it supports up to approximately 200 concurrent users per session -well above the typical class size of 50 students at DeVry University. Photon Fusion is a high- performance state transfer network SDK (software development kit), making it well suited for simulating a metaverse-based classroom learning environment where seamless interaction among students and instructors is essential. 

\setlength{\parskip}{1em}
The backend layer comprises servers, an analytics engine, and a secure database for tracking user performance, session logs (e.g. participation and engagement), quiz scores, and user feedback. Photon Enterprise Cloud is employed to host , operate , and scale the XR gamification applications developed for XR unity. It provides reliable servers and elastic scaling capabilities, which are critical for ensuring smooth operation- especially during the “Generative AI-XR scenarios” phase, which may demand high computational resources and dynamic server scaling . The system supports performance analytics for evaluating user interface (UX), including metrics from virtual classroom discussions, interactive circuit-building tasks, scenario-based troubleshooting exercises.  These analytics inform assessment of student engagement and learning outcomes.

\setlength{\parskip}{1em}
To implement adaptive and personalized learning with XR, a hybrid AI architecture can be employed, combining supervised learning for performance prediction and reinforcement learning integrated with generative AI for replicating XR scenarios optimized for the student learning curve. The model can continuously learn from user interaction data and iteratively refine the generated XR environments to enhance engagement and learning outcomes.

\section{Privacy and Security Framework in XR Gamification }
Abraham et al. (2022) highlighted that XR devices, being equipped with numerous embedded sensors, pose inherent cybersecurity risk . Some of the key challenges included: (i) amplification of biases through perception filters (moderate security risk), (ii) virtual threats that may result in physical harm ( extreme security risk), and (iii) lack of clear opt-out mechanisms for users ( low security risks) [30]. Warin and Reinhardt (2022) further emphasized that XR platforms collect significant volumes of biometric data via Natural User Interfaces (such as retina tracking for eye movement), which are also used in sensitive applications like banking authentication, thereby increasing vulnerability to cyber-attacks [31]. Additionally, the XR application’s operating system must be robustly designed to prevent malicious software, bugs, and unauthorized access, reducing the potential for cyber intrusion [32].
\setlength{\parskip}{1em}

\begin{table}[htbp]
\centering
\scriptsize
\caption{Cybersecurity risks and mitigations for XR-AI education platform}
\label{tab:cybersecurity}
\renewcommand{\arraystretch}{1.15}
\begin{tabular}{|p{1.5cm}|p{2.0cm}|p{0.5cm}|p{2.4cm}|p{0.7cm}|}
\hline
\textbf{Category} & \textbf{Threat} & \textbf{Sev.} & \textbf{Mitigation} & \textbf{Ref.} \\
\hline
Virtual harm & Disorientation / unsafe XR use & High & Env. calibration, safe zones, AI supervision & [36,37] \\
\hline
Biometric re-use & XR biometrics in fraud / ID theft & High & Role-based access, limited retention, user control & [38] \\
\hline
Opt-out limit & No choice to decline tracking & Low & Clear policy, opt-in/out toggle, anon. mode & [39] \\
\hline
Malware/virus & Exploited app vulnerabilities & High & Patching, secure OS, sandboxing, pen tests & [40,41] \\
\hline
Data sharing & Exposure of academic records & Mod. & FERPA APIs, access logging & [42,43] \\
\hline
Behav. surv. & Profiling or manipulation of users & Mod. & Diff. privacy, audits, limited monitoring & [44,45] \\
\hline
Sensor exploit & Unauthorized biometric access & High & End-to-end encryption, GDPR/FERPA compliance & [38,46] \\
\hline
Percept. manip. & Bias/filters distort XR content & Mod. & Algo. audits, inclusive data, overrides & [47] \\
\hline
\end{tabular}
\end{table}

\setlength{\parskip}{1em}
The risk categories outlined in Table-I necessitate a security-by-design approach where threat mitigation is embedded across all architectural layers. Some of the proposed mitigation strategies to address the cybersecurity and privacy risks inherent in the proposed XR-based educational systems include a multi-layered approach, combining technical, regulatory, and procedural safeguards. At the technical level, end-to-end encryption of biometric and interaction data should be strictly enforced both at the device and server layers. For the device layer, data transmission must be secured using industry standard protocols such as Transport Layer Security (TLS) 1.3 or through Virtual Private Network (VPN) tunnels, ensuring confidentiality and integrity during transit [33]. On the server side, sensitive data stored using Advanced Encryption Standards (AES) with 256-bit keys (AES-256), which is classified as military-grade encryption which complies with National Institute of Standard and Technology (NIST) and International Organization for Standardization (ISO) guidelines [34]. These encryption protocols not only protect against interception and unauthorized access but also support regulatory compliance with frameworks such as the General Data Protection Regulation (GDPR) and the Family Educational Rights and Privacy Act (FERPA). 

\setlength{\parskip}{1em}
Authentication protocols are also important components during assessment and ensuring integrity in the learning procedure. Authentication protocols such as multi-factor authentication (MFA) and role-based access control (RBAC) should be adopted to ensure that only authorized users ( such as educators, administration, and IT personnel) are allowed to modify and access sensitive information. By assigning permissions only necessary for each role, RBAC enforces the principle of least privilege, minimizing the attack surface and reducing the risk of internal misuse or accidental data leaks. For XR-AI platforms dealing with personal biometrics data and educational performance metrics, implementing both MFA and RBAC are critical to ensure compliance by educational regulatory bodies and to maintain data integrity and trust within the learning ecosystem. These protocols need to be integrated both at the device layer and backend layer, ensuring end-end access control and user accountability. 

\setlength{\parskip}{1em}
Privacy preserving features must be embedded within the XR platform, including the ability for users to anonymize data, opt out of data tracking , and control consent for data collection. In XR environments, where XR user interaction, behavior and biometric responses are continuously monitored for assessment, it is critical to provide mechanisms that uphold privacy rights. The XR platform should allow clear and accessible opt-out mechanism. Under such circumstances, alternate methods need to be employed to analyze student progress (instead of their biometric data). Another alternate approach includes (i) pseudonymization, which is a process where individual information is replaced with artificial identifiers and still allows scope for data analysis, (ii) data aggregation and (iii) masking [35]. 

\setlength{\parskip}{1em}
Building on these safeguards, the proposed XR-AI educational platform incorporates a security-by-design layered architecture, where each layer implements its own security protocols while maintaining interoperability with other layers. This defense-in-depth strategy ensures that even if one security layer is compromised, other layers continue to protect sensitive student information and preserve system functionality. These measures can be implemented in the architectural blueprint so that it supports regulatory compliance throughout system’s lifestyle. 

\setlength{\parskip}{1em}

\begin{figure}[htbp]
\centering
\includegraphics[width=0.9\linewidth]{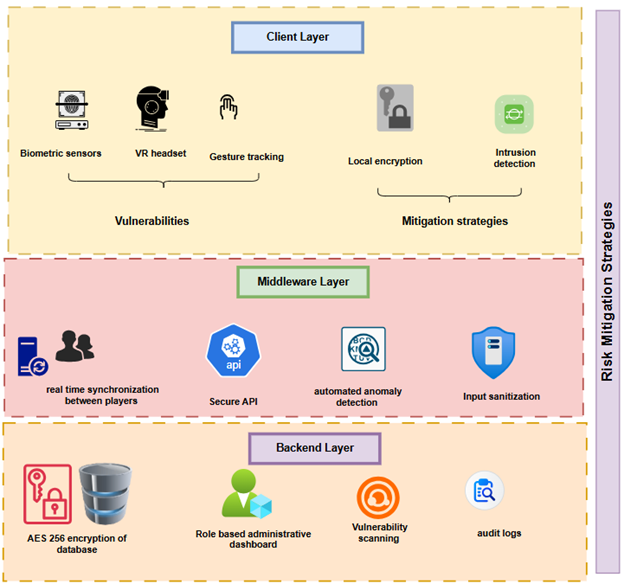} % scale down
\caption{Illustration of client-middleware-backend interaction and their risk mitigation strategies. 
}
\label{fig:Risk}
\end{figure}

\setlength{\parskip}{1em}
Fig. 5 shows the illustration of the client-middleware-backend interaction and their respective security mechanisms.  At the client layer, biometric sensors, gesture tracking devices, AR/VR headsets implement local encryption modules, device level intrusion detection, and safety zone calibration to prevent physical hazards.  The middleware layer handles real-time synchronization between XR clients, integrating secure APIs, input sanitization, and automated anomaly detection algorithms to identify irregular interaction patterns indicative of a security breach. The backend layer employs hardened  cloud infrastructure with AES 256 encrypted databases, role-restricted administrative dashboards, continuous vulnerability scanning , and immutable audit logs for incident forensics. 
\section{Summary and Future Directions }

This paper presents the design architecture of a dynamic gamification system within an Extended Reality- Artificial Intelligence (XR-AI) virtual training environment. The proposed architecture is implementable in Unity-XR or similar platforms, serving as a supplemental tool for STEM lessons that traditionally require in-person training or kinesthetic learning approaches to real-world case studies. The primary advantage  of an AI led XR based gamification system is its adaptivity, allowing virtual learning environment to dynamically adjust based on student performance, engagement levels, tokens earned and assessment scores.

\setlength{\parskip}{1em}
The architecture is structured in four sequential phases: (i) Introduction phase, (ii) Component Development phase, (iii) Fault introduction  and Correction phase, and (iv) Generative AI-XR scenarios phase. The Introduction phase imports a virtual classroom for DeVry University using the Unity Assets Marketplace, as illustrated in Fig.2 . In the Component Development Phase, basic 3D CAD models  of electrical components- such as resistors, capacitors, and inductors sourced from Sketchfab. The custom C\# script is developed for each component’s behavior, which undergo initial testing prior to deployment. The third and fourth phases require AI integration within  Unity to enable adaptive learning pathways and scenario generation based on learner profiles. 

\setlength{\parskip}{1em}
Given the sensitive nature of biometric and interaction data in XR environments, privacy and security are integral to the client-middleware-backend architecture. As summarized in Fig., each layer implements independent yet interoperable security controls including AES 256 encryption, role-based access control (RBAC), multifactor authentication (MFA), and safety zone calibration This defense-in-depth strategy ensures that even if one security layer is compromised, regulatory compliance with the General Data Protection Regulation (GDPR) and the Family Educational Rights and Privacy Act (FERPA) is preserved.

\setlength{\parskip}{1em}
With advancement of smart cities, such XR-AI metaverse classrooms can act as a systematic framework for next generation education [48-50]. While XR platforms are already deployed in medical and industrial training, the proposed adaptive virtual laboratory extends these benefits to both undergraduate/ graduate programs and K-12 classrooms- particularly for complex physics and engineering concepts that benefit from visual and interactive demonstrations. Additionally,  the system’s AI assisted adaptivity may help address educator shortages in rural regions by high quality remote training. 

\setlength{\parskip}{1em}
However, widespread adoption presents several challenges: (i) development of XR gamified teaching resources requires significant time and specialized skills,  (ii) high initial training costs, especially in K-12 level environments with high staff turnover, (iii) potential reduction of educator oversight  in the Generative AI-XR scenarios phase,  (vi) shortage of advanced cybersecurity personnel in underfunded public institutions, and (v) upfront investment in VR headsets, high speed internet, and supervised virtual access. Addressing these limitations will require targeted funding strategies, educator training programs, and modular deployment models to ensure equitable access and sustainable operation. 

\setlength{\parskip}{1em}
In summary, the proposed XR-AI gamified virtual laboratory provides a technically robust, pedagogically adaptive, and security conscious framework for immersive education. The phased deployment strategy enables controlled scalability while the integration of AI-driven adaptivity and multi-layered security measures supports both engagement and compliance.

\vspace{12pt}

\end{document}